\documentclass[12pt]{article}

\usepackage[super]{natbib} 
\usepackage{color} 
\usepackage[table]{xcolor} 
\usepackage{paralist}
\usepackage{geometry}
\usepackage{graphicx,graphics}
\usepackage{amsmath}	
\usepackage{amssymb}	
\usepackage{newtxtext,newtxmath}
\usepackage[T1]{fontenc}
\usepackage{ae,aecompl}
\usepackage{epsfig}   
\usepackage{microtype}
\usepackage[utf8]{inputenc}
\usepackage{enumitem}
\usepackage{ragged2e}
\usepackage{times}
\newlist{thematic}{itemize}{8}
\setlist[thematic]{label=$\square$}
\usepackage{pifont}
\usepackage{enumitem} 
\usepackage{hyperref} 
\hypersetup{
    colorlinks,
    linkcolor={blue!80!black},
    citecolor={blue!80!black},
    urlcolor={blue!80!black}
}

\usepackage{bm}

\newcommand{\arcsec}{$^{\prime\prime}$}

\newcommand{\specialcell}[2][c]{%
  \begin{tabular}[#1]{@{}c@{}}#2\end{tabular}}

\def\capfontcom{\normalsize\color{gray!60!black}}
\long\def\@makecaption#1#2{%
  \vskip\abovecaptionskip
  {\centering
    \begin{minipage}{0.999\linewidth}
      \sbox\@tempboxa{{\sffamily\bfseries\small #1}\capfontcom\,--- #2}%
      \ifdim \wd\@tempboxa >\hsize
      {{\sffamily\bfseries\small #1}\capfontcom\,--- #2}
      \else
      \global \@minipagefalse
      \hb@xt@\hsize{\hfil\box\@tempboxa\hfil}%
      \fi
      \end{minipage}\par
      }
  \vskip\belowcaptionskip}

\definecolor{tablealt}{rgb}{0.77,0.85,1.0}

\def\aj{{AJ}}                   
\def\araa{{ARA\&A}}          
\def\apj{{ApJ}}                 
\def\apjl{{ApJ}}                
\def\apjs{ {ApJS}}

\def\mnras{ {MNRAS}}
\def\nar{ {New Astr. Rev.}}

\def\nat{ {Nature}}

\def\procspie{ {Proc.~SPIE}}

\newcommand{\be}{\begin{equation}}
\newcommand{\ee}{\end{equation}}

\newcommand{\gtsima}{$\; \buildrel > \over \sim \;$}
\newcommand{\ltsima}{$\; \buildrel < \over \sim \;$}
\newcommand{\prosima}{$\; \buildrel \propto \over \sim \;$}
\newcommand{\gsim}{\lower.5ex\hbox{\gtsima}}
\newcommand{\lsim}{\lower.5ex\hbox{\ltsima}}
\newcommand{\simgt}{\lower.5ex\hbox{\gtsima}}
\newcommand{\simlt}{\lower.5ex\hbox{\ltsima}}
\newcommand{\simpr}{\lower.5ex\hbox{\prosima}}

\newcommand{\arcmin}{$^{\prime}$}

\begin{document}

\pagenumbering{roman}

\raggedright
\huge
Astro2020 Science White Paper \linebreak

Hot Drivers of Stellar Feedback from 10 to 10,000~pc 
\linebreak

\normalsize

\noindent \textbf{Thematic Areas:} \hspace*{60pt} $\square$ Planetary Systems \hspace*{10pt} $\boxtimes$ Star and Planet Formation \hspace*{20pt}\linebreak
$\square$ Formation and Evolution of Compact Objects \hspace*{31pt} $\square$ Cosmology and Fundamental Physics \linebreak
  $\boxtimes$  Stars and Stellar Evolution \hspace*{1pt} $\square$ Resolved Stellar Populations and their Environments \hspace*{40pt} \linebreak
  $\boxtimes$    Galaxy Evolution   \hspace*{45pt} $\square$             Multi-Messenger Astronomy and Astrophysics \hspace*{65pt} \linebreak
  
\textbf{Principal Author:}

Name: Edmund Hodges-Kluck
 \linebreak						
Institution: University of Maryland/NASA GSFC
 \linebreak
Email: edmund.hodges-kluck@nasa.gov
 \linebreak
 
\textbf{Co-authors:}  \linebreak
Laura A. Lopez (The Ohio State University), Mihoko Yukita (Johns Hopkins University/NASA GSFC), Andrew Ptak (NASA GSFC), Douglas Swartz (NASA MSFC/USRA), Panayiotis Tzanavaris (University of Maryland, Baltimore County/NASA GSFC), Sylvain Veilleux (University of Maryland), Joel Bregman (University of Michigan) \\

\justify

\textbf{Abstract:} Stellar feedback -- stars regulating further star formation through the injection of energy and momentum into the interstellar medium -- operates through a complex set of processes that originate in star clusters but shape entire galaxies. A mature theory of stellar feedback is essential to a complete theory of star and galaxy formation, but the energy and momentum injected by hot gas into its surroundings remains unclear. With a next-generation X-ray observatory, breakthrough progress can be made through precision measurements of the temperature, density, velocity, and abundances of hot gas on scales of star clusters to galactic superwinds.

\pagebreak

\setcounter{page}{1}
\pagenumbering{arabic}

\noindent \textbf{\sf \large 1. The Importance of Stellar Feedback -- and its Uncertainties}
\medskip

Stellar feedback -- the injection of energy and momentum by stars into their environment -- originates at the small scales of star clusters ($<$10~pc), yet shapes galaxies on large scales ($>$10~kpc); it is necessary for forming realistic galaxies in simulations\cite{White78,Hopkins2014} and accounting for observed galaxy properties\cite{Kennicutt1998,Tremonti2004,Muratov2015}. Stellar feedback maintains the multi-phase structure of the interstellar medium (ISM)\cite{McKee77}, inefficient star formation\cite{Zuckerman74,Krumholz07}, and the life-cycle of giant molecular clouds (GMCs)\cite{matzner02,Krumholz06}. Meanwhile, it generates hot gas, radiation, and cosmic rays that can expel gas from galaxies\cite{Veilleux2005,Heckman2017}.

The nature of stellar feedback is the source of some of the largest uncertainties in star- and galaxy-formation models\cite{dobbs06,Hopkins2014}. For example, depending on how feedback is implemented, the properties of Milky Way-like galaxies at $z = 0$ can vary by orders of magnitude\cite{Kim2014,Scannapieco2012}. These uncertainties stem primarily from four challenges: 1) the large dynamic range over which feedback is important; 2) the need to consider multiple modes of interconnected feedback; 3) the question of where energy and momentum are deposited; and 4) a dearth of observational constraints. 

Recent ``zoom-in'' simulations on parsec scales\cite{Agertz13,Renaud13,Ceverino14,Hopkins2014} are starting to overcome some of these issues by incorporating more relevant sub-grid physics. However, observations are essential for testing and improving these models. In particular, the role of hot ($\sim$10$^{7}$-$10^{8}$~K) gas shock-heated by stellar winds and supernovae (SNe) remains unclear, on both the small scales of H~{\sc ii} regions and on the large scales of galactic winds. Precise measurements of the energy, momentum, distribution, and metal content of the hot gas over four orders of magnitude in scale are needed to address these outstanding questions, and motivate the need for next-generation X-ray missions.

\bigskip
\noindent \textbf{\large \sf 2. Feedback in Star Clusters ($<$10~pc)}
\medskip

Fast stellar winds and SNe carve out large cavities, called superbubbles, that sweep up material from the surrounding medium. Further, SNe within the bubble thermalize efficiently, and the bubble becomes filled with tenuous, hot ($\sim10^{7}$~K), shock-heated gas\cite{castor75,weaver77,chu90,silich2005,rogers14}. As the bubble expands, much of the energy is radiated away, but the remainder can disrupt cool clouds and inject turbulence into the ISM\cite{McKee77,Hopkins2012}. This basic picture has been validated by observations: hot gas in bubbles detected by \textit{Chandra} and \textit{XMM-Newton}\cite{townsley06,gudel08} is consistent with models for superbubble expansion\cite{harperclark09,wareing18}. 

To take the next step and implement realistic superbubble models into feedback simulations requires understanding how superbubbles grow in different environments, and how the hot gas depends on the mass, metallicity, and ambient ISM density of the natal star cluster. These are not minor problems: studies of H~{\sc ii} regions in the Milky Way and LMC find temperatures between $kT \approx 0.1$-$0.8$~keV ($T \approx 10^6$-$10^7$~K) and luminosities $L_{\rm X} \sim 10^{31}$-$10^{35}$~erg~s$^{-1}$\hspace{1pt}\cite{chu95,oey96,jaskot11}. The large dispersion in $L_{\rm X}$ reflects different severity of energy losses, e.g., from radiative and adiabatic cooling (depending on how much work the hot gas does on its surroundings), thermal conduction, dust heating, and leakage of hot gas through the H~{\sc ii} shells\cite{rosen14}. 

It is therefore insufficient to study just a few nearby H~{\sc ii} regions in great detail. Instead, we need an inventory of hot gas in thousands of H~{\sc ii} regions in the Local Group, providing $L_{\rm X}$, $kT$, and $Z/Z_{\odot}$ (or $Z_{\rm Fe}$ and $Z_{\rm O}$). This is not feasible with \textit{Chandra} or \textit{XMM-Newton}. The main emission lines from gas at $10^6$-$10^7$~K occur between 0.5-1~keV, where a collecting area $A_{\text{eff}} > 5,000$~cm$^2$ is needed to detect X-ray faint ($L_{\rm X} \sim 10^{32}$~erg~s$^{-1}$) superbubbles out to 1~Mpc (in $<$1~Ms exposures). Several hundred X-ray photons are needed to make basic measurements, so a large collecting area is essential. These requirements are met by \textit{Athena}\cite{Barret2016}. With its higher angular resolution ($\theta \lesssim 0.5^{\prime\prime}$), the \textit{Lynx} strategic mission\cite{Gaskin2018} would extend this study to gas leaking out of bubbles.

\begin{figure}
  \begin{center}
    \begin{minipage}[b]{0.62\linewidth}
      \raisebox{0.0cm}{\epsfig{file=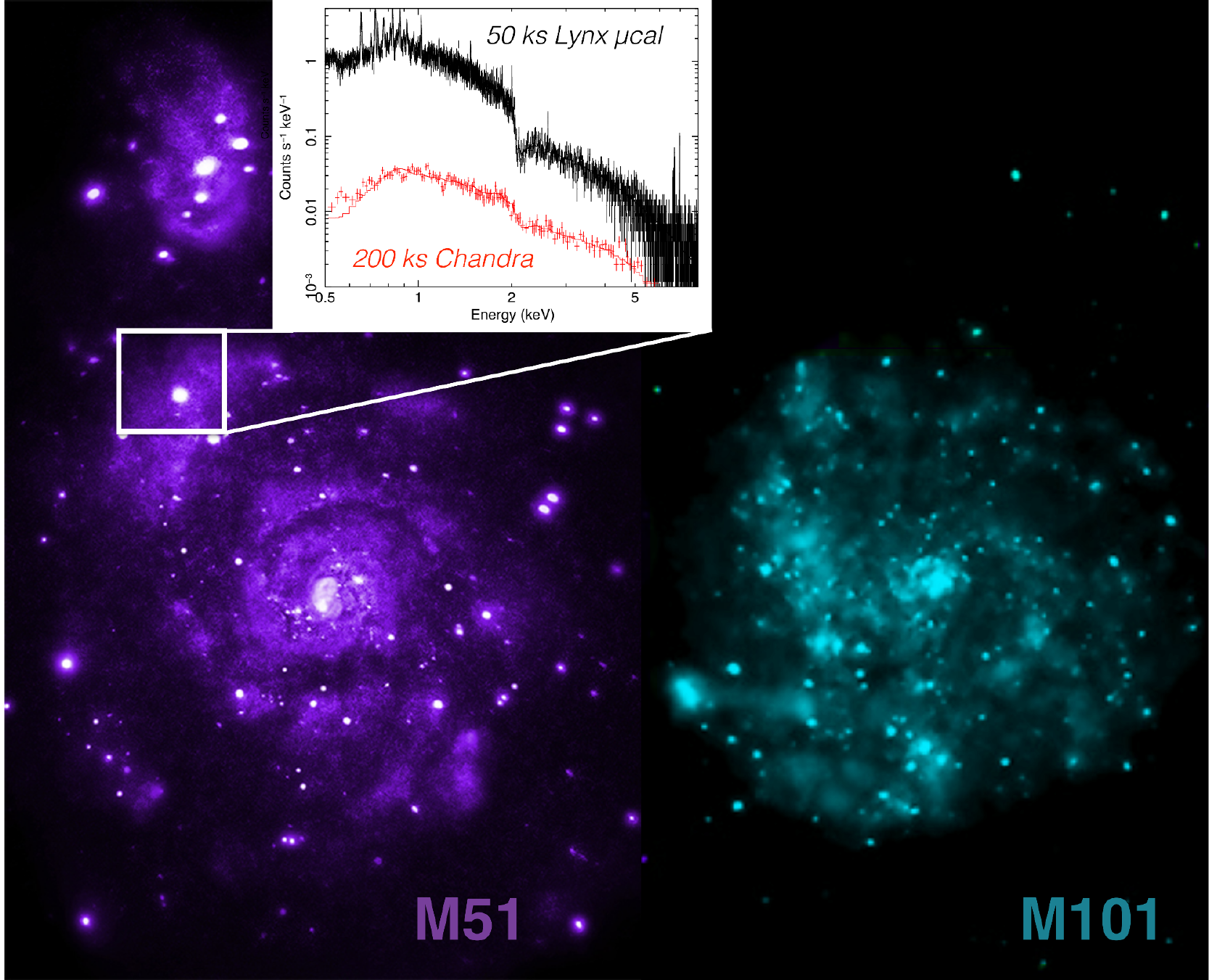, width=\linewidth}}
    \end{minipage}\hfill
    \begin{minipage}[b]{0.32\linewidth}
          \caption{\small {\it Chandra} images of M51 (left; 850~ks\cite{kuntz16}) and M101 (right; 1 Ms\cite{kuntz10}). The inset shows the 200~ks {\it Chandra} spectrum (red) of the diffuse gas from a 1\arcmin\ portion of the M51 disk. For comparison, a simulated spectrum (black, based on two thermal plasma models) from a 50-ks observation with the {\it Lynx} microcalorimeter is shown, which resolves emission lines from heavy elements (e.g., Si, Fe) in the ISM of M51. $\sim$1$^{\prime\prime}$ spatial resolution across the entire galaxy is needed to detect and mask point sources, and thus obtain an accurate inventory of the hot gas energy. \label{figure:diffusegas}}
    \end{minipage}
  \end{center}
\vskip-15pt
\end{figure}

Detecting brighter superbubbles with $L_{\rm X} = 10^{33}$-$10^{35}$~erg~s$^{-1}$ in external galaxies requires high angular resolution ($\theta < 1$\arcsec, corresponding to 25~pc and 50~pc at 5~Mpc and 10~Mpc). Figure~\ref{figure:diffusegas} shows what the hot ISM looks like with high resolution \textit{Chandra} observations\cite{kuntz10,kuntz16}; superbubbles appear as faint knots (which are hard to distinguish visibly from the brighter X-ray binaries). These luminosities also require high sensitivity ($A_{\text{eff}} > 5,000$~cm$^2$ for a calorimeter) and a wide field of view ($>$15~arcmin) to efficiently accumulate a sample of hundreds to thousands of superbubbles.

\bigskip
\noindent \textbf{\large \sf 3. Supernova Feedback from Star Clusters to Galaxies (10-1000 pc)}
\medskip

Hot gas in superbubbles compresses cooler gas, drives turbulence in the ISM, distributes metals, expels hot gas from the disk in fountain flows, and forms channels through which ionizing photons from young star clusters can escape the galaxy. These processes work in tandem with, and depend on, momentum injected by radiation pressure and cosmic rays. The energy and momentum of the hot ISM also depends on density, which controls radiative losses. Sophisticated simulations are needed to understand what happens after gas escapes from its home cluster.

Models are becoming less phenomenological and more physical: they include recipes for plasma heating and cooling, mixing of hot and cold gas, magnetic fields, stellar mass-loss, etc\cite{Dale2015}. However, the observational constraints on the hot ISM needed to inform these models remain poor. Important measurements include the thermalization efficiency of SNe ($\epsilon$) as a function of environment\cite{Hopkins2013,Vasiliev2015,Fierlinger2016}, the mass-loading factor ($\beta$) in outflows\cite{Dong_Zhang2014,Kim2018}, and the mass in hot gas as a function of temperature and density (a phase diagram). $\epsilon$ and $\beta$ characterize the quantitative impact of outflows and momentum injection into the ISM, while the phase diagram globally constrains ISM models. 

These quantities map to the temperature ($kT$) and density ($n$) of the hot plasma; for example, the temperature is proportional to the ratio $\epsilon/\beta$, and $\beta$ can be constrained with the radius of the hot region and the star-formation rate\cite{Chevalier1985,Strickland2009a,Zhang2014}. These measurements are challenging to make because the X-ray luminosity is proportional to $n^2$ (requiring much longer exposures to measure $n$ in tenuous regions, such as between spiral arms), and much of the X-ray emitting gas is not in collisional ionization equilibrium\cite{deAvillez2012}, so commonly used thermal plasma models are misleading.

\begin{figure}
    \centering
    \includegraphics[width=0.97\textwidth]{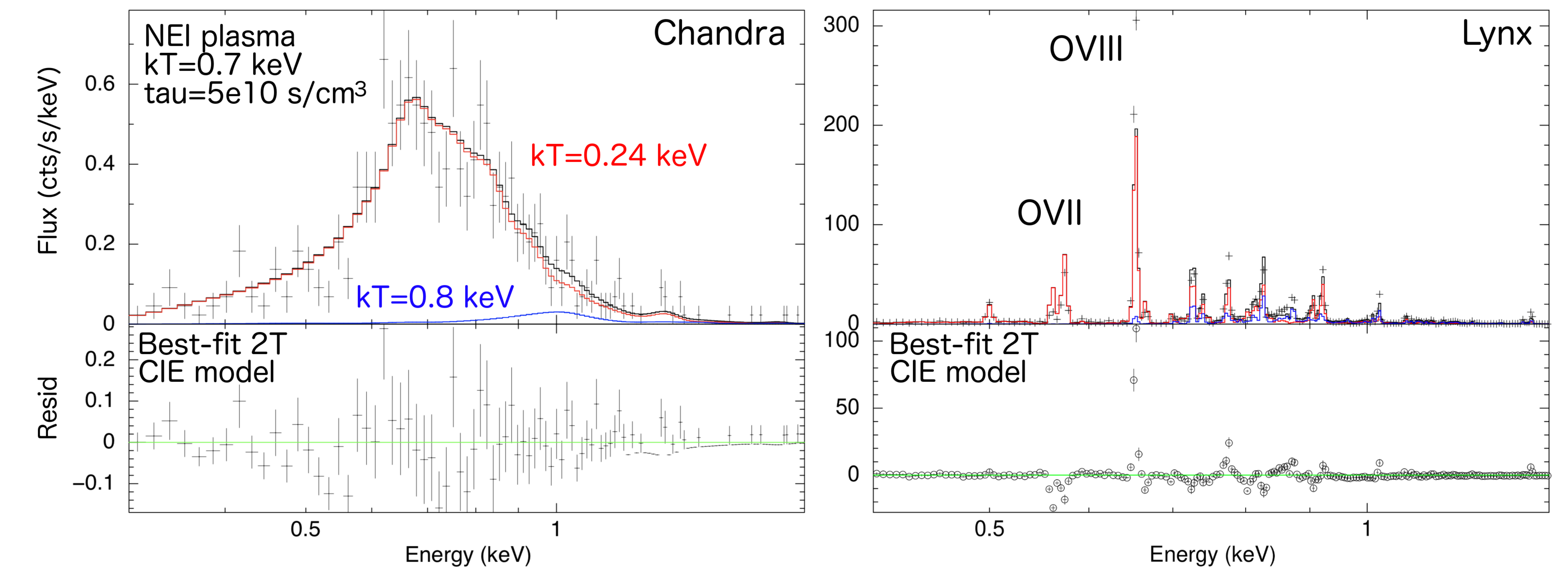}
    \vspace{-0.25cm}
    \caption{\small Much of the X-ray bright ISM may not be in collisional ionization equilibrium (CIE), but high energy resolution is needed to properly model it. \textbf{Left:} A simulated non-equilibrium ionization (NEI) plasma model at CCD energy resolution (with $kT=0.7$~keV and ionization timescale $\tau = 5\times 10^{10}$~s~cm$^{-3}$) is adequately fit by a CIE model with $kT_1 = 0.24$~keV and $kT_2 = 0.8$~keV. \textbf{Right:} When the main emission lines are resolved, no CIE model can possibly fit the data, as evidenced by a poor fit to the expected CIE line strengths.  \label{figure:NEI}}
    \vspace{-0.2cm}
\end{figure}

Current observational data are completely unmatched to the scope of the problem. Outside of the Local Group, few galaxies have enough signal from hot gas to make any measurements on the spatial scales relevant to constraining simulations, and instead the hot ISM properties are inferred from a single spectrum from the whole galaxy. Even in the few cases with deep {\it Chandra} data (Figure~\ref{figure:diffusegas}), measurements on sub-kpc scales are hindered by low energy resolution, which makes it difficult to distinguish non-equilibrium and collisional equilibrium models (Figure~\ref{figure:NEI}). Indeed, this may be why the hot ISM in most star-forming galaxies can be fit by a two-temperature collisional plasma\cite{Strickland2004,kuntz10,Li2013} with $kT_1 \sim 0.1$-$0.2$~keV and $kT_2 \sim 0.5$-$0.8$~keV. 

Accurate modeling of the hot ISM spectrum on scales of 0.1-1~kpc requires high angular resolution ($\theta < 1-10$\arcsec), a large collecting area ($A_{\text{eff}} > 2,000$~cm$^2$ at 1~keV) and energy resolution $\Delta E < 4$~eV. The \textit{Athena} X-IFU meets these requirements for nearby galaxies. $\theta < 1-2$\arcsec) is needed to fully inventory the hot gas in galaxies within 30~Mpc. 

Higher resolution ($\theta \lesssim 1^{\prime\prime}$) is needed to resolve the \textit{interfaces} of hot and cool gas in detail beyond the Local Group. \textit{Chandra} images reveal edges in the hot ISM related to stellar structure\cite{kuntz16}, magnetic fields, and filamentary structures\cite{Wang2001} seen in other wavelengths, and studies of galaxy clusters and supernova remnants highlight the need to resolve edges to capture the essential physics (see the white paper by Markevitch et al.). This is generally feasible only with \textit{Lynx}, although \textit{Athena}, in tandem with existing \textit{Chandra} data, will be able to study some nearby edges.

\bigskip
\noindent \textbf{\sf \large 4. What Drives Galactic Winds? (100-10,000~pc)}
\medskip

Efficient thermalization of SNe in starburst galaxies forms pockets of superheated ($T>10^8$~K) gas that break out of the disk and drive winds with hot component velocities $v_{\text{hot}}>1000$~km~s$^{-1}$ \hspace{2pt}\cite{Chevalier1985,Fielding2017}. The hot wind contains more than 90\% of the total wind energy and most of the metals\cite{Strickland2000}, but it may not be able to accelerate cool clouds (which dominate the wind mass) to the observed $v \lesssim 1000$~km~s$^{-1}$ without shredding them, so alternative momentum sources such as cosmic rays and radiation pressure have been explored\cite{Zhang2018}. The question remains, \textit{does} the hot wind couple to the mass (and \textit{how}?) How does the wind evolve as it expands, and how many metals does it carry? 

\begin{figure}
    \centering
    \includegraphics[width=0.98\textwidth]{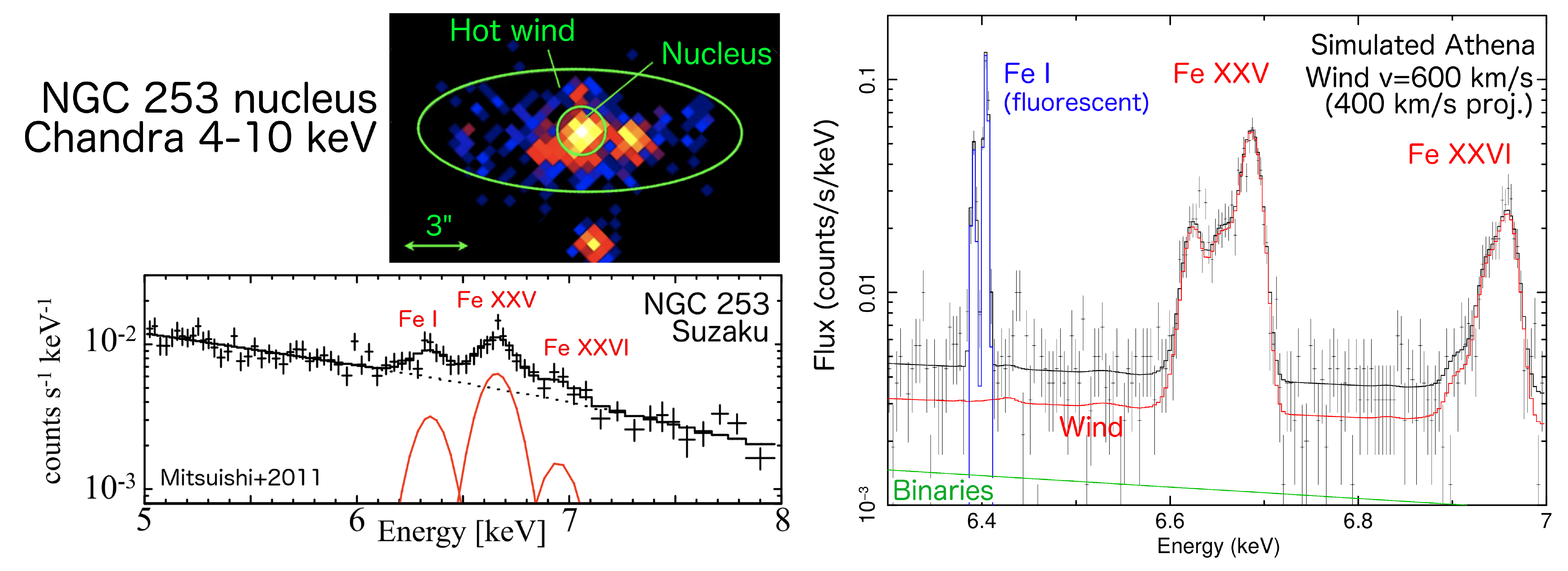}
    \vspace{-0.5cm}
    \caption{\small \textbf{Left}: \textit{Suzaku} detected diffuse hot gas in NGC~253 through Fe~{\sc xxv} and Fe~{\sc xxvi} lines\cite{Mitsuishi2011}, which however are unresolved, leading to a limit on $v_{\text{hot}}$. 
    \textbf{Right}: A simulated \textit{Athena} spectrum from the nucleus resolves the lines
    (Fe~I K$\alpha$ is fluorescent and not from the hot wind). \textbf{The input line-of-sight velocity is recovered by centroiding to} \bm{$\Delta v = 40$} \textbf{km~s}\bm{$^{-1}$} \textbf{(10\%)}, illustrating the power of a calorimeter.}
    \label{figure:hotwind}
    \vspace{-0.2cm}
\end{figure}

Understanding the hot wind begins with measuring its energy content, $E_{\text{wind}}$, which requires $n$, $kT$, and $v_{\text{hot}}$ at the wind's base. Superheated gas has been detected in the nearby superwind galaxies M82\cite{Strickland2009a,Liu2014} and NGC~253\cite{Mitsuishi2011} through diffuse Fe~{\sc xxv} emission (Figure~\ref{figure:hotwind}), but the energy resolutions ($\Delta E$) of existing spectrometers are too low to directly measure $v_{\text{hot}}$. To constrain $v_{\text{hot}}$ to a precision of 25\%, which limits the precision on $E_{\text{wind}}$, through line widths and centroids requires $\Delta E \leq 5$~eV at the 6.7~keV Fe~{\sc xxv} line (centroid accuracy is $\sim \Delta E/$[{\small $S/N$}]). The Resolve calorimeter on board \textit{XRISM}\footnote{The JAXA/NASA X-ray Imaging and Spectroscopy Mission\cite{Tashiro2018}, to replace \textit{Hitomi} in 2022.} ($\Delta E = 5$~eV, $\theta=1.6^{\prime}$) will do this for M82 and NGC~253. Line ratios within the Fe~{\sc xxv} and Fe~{\sc xxvi} complexes, using the same data, constrain $n$ and $kT$. 

The first measurements of $v_{\text{hot}}$ will be a breakthrough, but will not tell us how $E_{\text{wind}}$ relates to the star-formation rate, stellar mass, and metallicity. Strickland et al. (2009)\cite{Strickland2009b} identified about 25 winds sufficiently bright to observe with modest exposure times (100-200~ks) for $A_{\text{eff}} \sim 1000$-2000~cm$^2$. Isolating the superheated gas from bright X-ray binaries in the more distant galaxies requires angular resolution $\theta < 5$\arcsec-$10^{\prime\prime}$, while measuring $E_{\text{wind}}$ to 10\% requires $\Delta E \leq 2$~eV. 

A more ambitious goal that would connect the hot wind to its cooler surroundings (as measured with optical IFUs) involves \textit{mapping} the velocities of diffuse Fe~{\sc xxv} and Fe~{\sc xxvi}. Based on \textit{Chandra} images (e.g., Figure~\ref{figure:hotwind}), this would require higher angular resolution ($\theta < 1$\arcsec-$2$\arcsec) and correspondingly larger $A_{\text{eff}}$ for reasonable ($<$400~ks) exposures. Such measurements will be essential to understanding winds across cosmic time, as the underlying principles should be the same\cite{Heckman2017}.

The fate of the hot wind is unclear. At larger radii, winds emit soft X-rays from $10^{6-7}$~K gas, which may result from radiative cooling\cite{Thompson2016} or shocked material\cite{Veilleux2005}. At least some of the emission comes from charge exchange (3-87\% of the flux in strong lines\cite{Zhang2014,Cumbee2016}). Models make predictions\cite{Zhang2018} that can be tested with an X-ray calorimeter: in a radiatively cooling wind, $kT$ would drop rapidly with height and $v$ would drop gently\cite{Thompson2016}. On the other hand, shocked clouds would not (necessarily) decline in either temperature or velocity with radius. We need to measure the velocity of the $10^{6-7}$~K gas to 20-30\%, using lines such as Mg\,{\sc xvii} ($E=1.47$\,keV, $T_{\text{peak}}=10^7$\,K) and Ne\,{\sc x} ($E=1.02$\,keV, $T_{\text{peak}}=5\times 10^6$\,K). $n$ and $kT$ can be estimated from the He-like triplets in the 0.3--3~keV bandpass. These observations will also enable an inventory of the metal abundances in the hot gas. 

\textit{XRISM} can resolve several winds, but a factor of 10 better $\theta$ and $A_{\text{eff}}$ are needed for a sample\cite{Strickland2009b}. Even higher resolution is needed to move beyond phenomenology to understand the coupling of wind energy to cool gas in filaments, as soft X-ray winds are highly structured and may cool to form the warm filaments\cite{Thompson2016,Schneider2018}. Measuring velocities to 10-20\% in the $10^6$-$10^7$~keV gas and mapping abundances in individual filaments requires $\theta < 1^{\prime\prime}$, high spectral resolution ($\Delta E < 0.5$~eV, for lines at $E<1$~keV), and $A_{\text{eff}} > 10,000$~cm$^2$. X-ray absorption (i.e., using gratings) is also a promising way to study these winds. This rich topic is explored in the white paper by Tremblay et al.

\bigskip
\noindent \textbf{\sf \large 5. Winds and the Hot ISM out to $z\sim 1$}
\medskip

X-ray surveys to detect the most distant black holes are presented in other white papers, and the cameras needed to perform them (i.e., with $\theta \approx 1$\arcsec\ across a 15$^{\prime}$ field of view) can also detect bright winds -- and even the hot ISM -- at cosmological distances. ULIRG winds, such as in NGC~6240 or Arp~220, are bright enough to detect up to $z\sim 1$, if the wind can be resolved (1\arcsec\ at $z = 1.0$ corresponds to 8~kpc). The hot ISM is fainter, but could be resolved from X-ray binaries out to $z\sim 0.1$ in surveys, and to much larger distances ($z \sim 3$) in strongly lensed galaxies\cite{Dye2015}. The spectrum would recover the global temperature, density, and metallicity, subject to the caveats noted above.

\bigskip
\noindent \textbf{\sf \large 6. Instrumental Requirements}
\medskip

Here we summarize the capabilities needed to execute an ambitious program that will transform our understanding of stellar feedback from small ($<$10~pc) to large ($>$10,000~pc) scales. This will require high angular resolution coupled with high spectral resolution, such as what can be attained with a micro-calorimeter in a high throughput observatory. 
\bigskip

\hspace{-0.5cm}
\begin{tabular}{ |p{2in}|p{2in}|p{1.8in}|  }
\hline
\rowcolor{tablealt} \textbf{\sf Science Goals} & \textbf{\sf Methodology} & \textbf{\sf Required Capabilities} \\
\hline
\rowcolor{red!15!white}
Constrain how feedback properties
vary as a function of star cluster mass, galactic environment, and metallicity   &
Characterize hot gas in thousands of H{\sc ii} regions in the Local Group and in superbubbles up to 30 Mpc &
\specialcell[t]{A$_{\rm eff}$:  5,000 cm$^2$ @ 1 keV  \\ $\theta$: $<$10\arcsec (LG) \\ $\theta$: 0.5\arcsec - 1.0\arcsec \\ FoV: $\sim$15\arcmin x15\arcmin\ }
\\
\hline
\rowcolor{green!15!white}Constrain $\epsilon$ and $\beta$ on 100-2,000~pc scales and in different environments and galaxies &
Measure the temperature and density of hot gas in galaxies up to 30 Mpc &
\specialcell[t]{A$_{\rm eff}$:  2,000 cm$^2$ @ 1 keV  \\ $\theta$: 1\arcsec-10\arcsec \\ $\Delta E$: $<$4~eV}
\\
\hline
\rowcolor{cyan!20!white}Measure the velocity and energy of the hot wind &
Measure velocity of  hot wind via the 6.7 keV Fe~{\sc xxv} line in $\sim$25 nearby starbursts &
\specialcell[t]{A$_{\rm eff}$:  2,000 cm$^2$ @ 6 keV  \\ $\theta$: 10\arcsec \\ $\Delta E$: $<$2 eV}
\\
\cline{2-3}
\rowcolor{cyan!20!white}
&
Map the velocities of diffuse Fe~{\sc xxv} and Fe {\sc xxvi} &
\specialcell[t]{A$_{\rm eff}$:  3,000 cm$^2$ @ 6 keV  \\ $\theta$: 1-2\arcsec \\ $\Delta E$: $<$3-4 eV} \\
\hline
\rowcolor{cyan!20!white}
Understand the origin of the soft X-rays from galactic winds, and measure metal mass &
Measure the velocity, temperature, and abundances as a function of wind radius in 20-30 systems &
\specialcell[t]{A$_{\rm eff}$:  2,000 cm$^2$ @ 1 keV  \\ $\theta$: 3-20\arcsec \\ $\Delta E$: $<$2 eV \\ $\theta$: 0.5\arcsec\ (filaments) \\ $\Delta E$: 0.5~eV (filaments)}
\\
\hline
\rowcolor{blue!20!white}Measure hot gas in and around galaxies to $z\sim 1$ &
Measure Wind/ISM temperature, density, and metallicity  & 
\specialcell[t]{A$_{\rm eff}$:  5,000 cm$^2$ @ 0.5 keV  \\ $\theta$: $<$1\arcsec \\ FoV: $\sim$15\arcmin x15\arcmin\ } 
\\
\hline
\end{tabular}

\clearpage



\end{document}